\documentclass[aps,prl,reprint,superscriptaddress]{revtex4-2}

\usepackage{times}
\usepackage{scalerel}
\usepackage{amssymb}
\usepackage{suffix}
\usepackage{float}
\usepackage{mathtools}
\usepackage[utf8]{inputenc}
\usepackage{booktabs}
\usepackage{cases}
\usepackage[multiple]{footmisc}
\usepackage{dcolumn}
\usepackage{color,soul}
\usepackage{rotating}
\usepackage{perpage}
\usepackage{xcolor}
\usepackage{soul}
\usepackage{tikz}
\usepackage[T1]{fontenc}
\usepackage{etoolbox}
\usepackage{graphics}
\usepackage{siunitx}
\usepackage{hyperref}
\usepackage{float}	
\usepackage{collref}
\usepackage{multirow}
\usepackage{mathtools}
\usepackage{bm}
\usepackage{url}

\usepackage{tikz}
\usepackage{tikz-3dplot}
\usepackage{changes}

\newcommand{\blue}[1]{\textcolor{blue}{#1}}

\begin{document}


\title{Ultrafast dynamics of a fermion chain in a terahertz field-driven optical cavity}






	\author{Mohsen Yarmohammadi}
	\email{mohsen.yarmohammadi@utdallas.edu}
	\address{Department of Physics, The University of Texas at Dallas, Richardson, Texas 75080, USA}
 \author{John Sous}
	\email{jsous@ucsd.edu}
	\affiliation{Department of Chemistry and Biochemistry, University of California, San Diego, La Jolla, California 92093, USA}
	\author{Marin Bukov}   
	\email{mgbukov@pks.mpg.de}
	\affiliation{Max Planck Institute for the Physics of Complex Systems, N\"othnitzer Str.~38, 01187 Dresden, Germany}
	\author{Michael H. Kolodrubetz}
	\email{mkolodru@utdallas.edu}
	\address{Department of Physics, The University of Texas at Dallas, Richardson, Texas 75080, USA}
	\date{\today}
	
	\begin{abstract}
		We study the effect of a terahertz field-driven single cavity mode for ultrafast control of a fermion chain with dissipation-induced nonlinearity and quadratic coupling to an infrared-active phonon mode. Without photon loss from the cavity, we uncover a first-order phase transition in the nonequilibrium steady state only for the lower phonon-polariton, accompanied by polaritons whose frequency response is asymmetric with respect to the photon frequency due to the direct laser-induced dressing effect on the photon. A weak laser field fails to induce the phase transition but renders the polaritons symmetrical. Finally, we show that sufficiently strong photon loss from the cavity eliminates the polaritons and the associated phase transition. The experimental feasibility of these phenomena is also proposed.
	\end{abstract}
	\maketitle
	{\allowdisplaybreaks
 
		\textbf{\blue{\textit{Introduction.}}}---Laser fields~\cite{Forst2011,Mitrano2016,PhysRevB.94.214504,PhysRevLett.118.087002,Kennes2017,PhysRevB.95.205111} may provide control over quantum materials, possibly revealing phenomena such as light-induced topology~\cite{Claassen2016,Hubener2017}, laser-induced magnetism~\cite{Shin2018,yarmohammadi2023laserenhanced}, ferroelectric phase-transitions~\cite{PhysRevX.10.041027}, and superconductivity~\cite{thomas2019exploring}. Classical light fields often introduce challenges associated with heating which may be circumvented via the use of optical cavities~\cite{Ruggenthaler2018,FriskKockum2019,Hubener2021,10.1063/PT.3.4749} to extend the lifetime of laser-induced states. Placing laser-driven systems in optical cavities not only establishes steady states but also extends their preservation by suppressing disruptive heating effects~\cite{Ruggenthaler2018,FriskKockum2019,Hubener2021,10.1063/PT.3.4749,10.1063/5.0083825,LeDe_2022}. The effectiveness of light-matter coupling in terahertz cavities, where quantum fluctuations play a substantial role in influencing the matter degrees of freedom, is of great importance for tuning the lifetime of nonequilibrium states~\cite{Li2018}.
            
        Understanding steady-state behavior requires systematic consideration of dissipation~\cite{PhysRevB.107.174415,PhysRevB.108.L140305,yarmohammadi2023laserenhanced,yarmohammadi2020dynamical}, affecting the balance between injected laser energy and energy loss to the environment. This interplay facilitates the emergence of intriguing nonthermal states of matter. In a recent study~\cite{PhysRevB.108.L140305}, we uncovered a novel first-order phase transition in the steady state of a fermion chain coupled to laser-driven infrared~(IR)-active phonons. This transition arises from the dressed dispersion of electrons in dissipation processes. For practical applications in laser-driven switching and control processes~\cite{7893184,PhysRevB.107.024413,doi:10.1080/01411594.2014.971322}, tuning the timescale of the transient process for achieving nonequilibrium steady states (NESS) becomes crucial. In light of the significant role of the cavity in adjusting the timescale of steady states in a driven-dissipative material, placing the above fermion chain into an optical cavity~\cite{PhysRevLett.122.017401} emerges as a promising approach to engineering the system's dynamic responses. 
           
        This Letter addresses three questions. First, with the inclusion of laser-induced transient excitations into the dissipation process in a fermion chain~\cite{PhysRevB.108.L140305}, how can we manipulate the NESS by coupling electrons and phonons to a single cavity mode? Second, does the observed first-order phase transition in the cavity-free model persist in the coupled cavity-fermion chain, or do we observe different phase transitions? Lastly, what effect does photon loss from the cavity have on the dynamic characteristics of the system? In answering these questions, we discover a first-order phase transition only specific to lower phonon-polaritons, linked to photon's asymmetric responses, a consequence of dressing effects induced by the laser. Our findings reveal that a strong photon loss from the cavity destroys both the phase transition and polaritons. 
            

            \textbf{\blue{\textit{Model.}}}---The model, Fig.~\ref{f1}, comprises of seven terms, including spinless electrons (assuming spin-polarized bands for simplicity), phonons, a single photon mode, electron-phonon coupling, phonon-photon coupling, electron-photon coupling, and the interaction between the laser and the single photon mode~\cite{LeDe_2022,Eckhardt2022,yarmohammadi2020dynamical,PhysRevB.108.L140305,mahan2013many}. We employ the dipole approximation for long-wavelength optical cavities, neglecting the spatial dependence of the electromagnetic vector potential~\cite{LeDe_2022,Eckhardt2022}. 
            
            The total time-dependent Hamiltonian (we set $\hbar = 1$) is:\begin{equation}\label{eq_1}
            	\begin{aligned}
            		&\mathcal{H}(t)  =  -t_0 \sum_\ell \big(c^\dagger_\ell c_{\ell+1} + {\rm H.c.}\big) + \omega_0 \sum_\ell a^\dagger_\ell a_\ell  \\ 
            		&+ \eta b^\dagger b + g_{\rm q}  \sum_\ell \big(a^\dagger_\ell + a_\ell\big)^2 \Big(c^\dagger_\ell c_\ell -\frac{1}{2}\Big) \\
            		& -\frac{i\Delta}{\sqrt{L}} (b^\dagger  + b ) \sum_\ell (a^\dagger_\ell -a_\ell) + \frac{2\Delta^2}{L \omega_0} (b^\dagger  + b )^2\\
            		& -t_0 \sum_\ell \big(e^{-i\frac{\lambda}{\sqrt{L}} (b^\dagger+ b)} c^\dagger_\ell c_{\ell+1}  + {\rm H.c.}\big) +\frac{\mathcal{E}(t) }{\sqrt{L}}\big(b^\dagger + b \big)\, ,
            	\end{aligned}
            \end{equation}which incorporates the electronic nearest-neighbor hopping amplitude $t_0$, optical phonon ($\omega_0$), and photon ($\eta$) frequencies. Operators $c^\dagger_\ell~(c_\ell)$, $a^\dagger_\ell~(a_\ell)$, and $b^\dagger~(b)$ create~(annihilate) electrons, phonons, and photon at site $\ell \in \{1,\dots,L\}$ in a chain of length $L$. Quadratic electron-phonon coupling is denoted by $g_{\rm q}$, and $\lambda$ represents the dimensionless electron-photon coupling strength. $\Delta = \alpha\,\eta\sqrt{\omega_0/8\eta}$ characterizes the phonon-photon coupling with $\alpha \approx 0.02$ to 0.2 being the polariton-to-photon frequency ratio in our simulations~\cite{LeDe_2022,Eckhardt2022,mahan2013many}. To coherently drive the cavity, we use a steady laser field $\mathcal{E}(t) = \mathcal{A}_0 \cos(\omega \,t)$ with frequency $\omega$ and amplitude $\mathcal{A}_0$.\begin{figure}[t]
		 	\centering
		 	\setlength\abovecaptionskip{1pt}
		 	\includegraphics[width=0.85\linewidth]{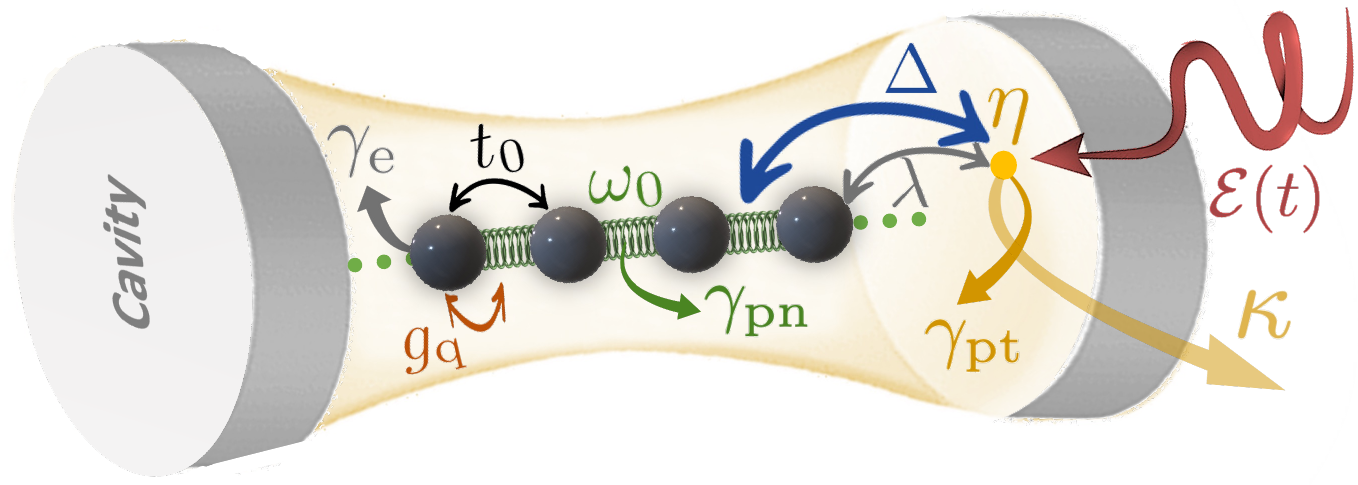}
		 	\caption{\textbf{A fermion chain in a laser-driven optical cavity.} The model depicts a driven cavity-fermion chain with a laser described as a classical field $\mathcal{E}(t)$, driving a resonant single photon mode $\eta$ in the cavity. The chain includes electrons that hop with energy $t_0$ and optical IR-active phonons $\omega_0$, where $g_{\rm q}$ denotes quadratic electron-phonon coupling. The electron-photon~(phonon-photon) coupling strength is denoted by the parameter $\lambda$~($\Delta$). Damping of the driven photon, phonon, and electron to the internal bath~(a combination of other phonons and an electromagnetic field of the non-driven photons) is represented by the rates $\gamma_{\rm pt}$, $\gamma_{\rm pn}$, and $\gamma_{\rm e}$, respectively. Additionally, the damping of the photon to the external bath (vacuum) from the cavity is defined by $\kappa$.}
		 	\label{f1}
		 \end{figure}

            Our model serves as a minimal descriptor of the ultrafast dynamics induced by dissipation and nonlinear electron-phonon coupling in a terahertz field-driven cavity and may be relevant to materials with symmetry constraints on their electron-phonon coupling if placed in a cavity, such as ET-F$_2$TCNQ, (Pb,Bi)$_2$Sr$_2$CaCu$_2$O$_8$, YBa$_2$Cu$_3$O$_{6+x}$, K$_3$C$_{60}$, and $\kappa$-salt~\cite{Kaiser2014,PhysRevLett.115.187401,PhysRevLett.91.167002,Kennes2017,PhysRevB.95.205111,PhysRevB.98.165138,Sous2021,PhysRevX.10.031028}. Application of our model to these materials requires the inclusion of other material-specific details, which we leave to future work. 
            
           We study the dynamics for small values of the electron-photon coupling~\cite{Li2018} in compliance with the dipole approximation, first, and second, to have a negligible energy density effect on the electrons. We focus only on the zero-momentum mode of the dispersionless phonon $a_0$, as it is the only mode that couples to electrons or photons in our translation-invariant model under the dipole approximation. This comes from the fact that phonons are generated by atoms spaced by order of an \AA, while the laser wavelengths are on the order of nanometers. Since the relative quantum fluctuations in phonon occupation approach zero in the thermodynamic limit, the mean-field decoupling approximation is then employed for describing the interaction between particles~\cite{yarmohammadi2020dynamical}.
		
		  	To consider dissipation across all particles, we introduce an internal bath (a combination of other phonons of the chain and the electromagnetic field inside the cavity from non-driven cavity photon modes) and an external vacuum bath from the cavity to simulate photon loss. This is implemented using the Lindblad quantum master equation framework, incorporating phenomenological damping rates~\cite{lindblad1976,breuer2007theory}. The equation governing coherent evolution and the dissipator for any arbitrary observable $O$ is then expressed as $\langle \dot{O}\rangle (t) =   i\langle[\mathcal{H},O(t)]\rangle  +  \sum_{j} \Gamma_{j} \Big<\mathcal{L}_{j}^{\dagger}O(t)\mathcal{L}_{j}  -\frac{1}{2}\big\{\mathcal{L}_{j}^{\dagger}\mathcal{L}_{j},O(t)\big\}\Big>$. The Lindblad jump operators $\mathcal{L}_j= O$ and $O^\dagger$, with $O$ being any of $\{a,b,c\}$ operators~\cite{PhysRevB.108.L140305}, have corresponding damping parameters $\Gamma_j = \gamma_{\rm pn}$, $\gamma_{\rm pt}$, and $\gamma_{\rm e}$ for the phonon, photon, and electron subsystems inside the cavity. Additionally, $\Gamma_j = \kappa$ represents the damping parameter for cavity photon loss. 
  
        By incorporating laser-induced transient excitations into the dissipation processes, the ratio between the damping of two dissipators in the electronic subsystem results in a modulated dispersion $\widetilde{\varepsilon}_k(t)$~\cite{lindblad1976,breuer2007theory}. Modes $k$ in the vicinity of the Fermi level $k_{\rm F}$ play a key role in the laser modulation of electron occupation, influenced by electron-photon and electron-phonon couplings. Thus, by expanding the electronic dispersion around the Fermi level, we obtain\begin{equation}\label{eq_6}
        	\widetilde{\varepsilon}_k(t) \approx  - 2\, t_0 (k-k_{\rm F}) + g_{\rm q} q^2_{\rm pn}(t) + 2\, t_0\, \lambda\, q_{\rm pt}(t)\, ,
        \end{equation}where $q_{\rm pt}(t) =L^{-1/2} \langle b^\dagger  + b  \rangle (t)$ and $q_{\rm pn}(t) =L^{-1/2}  \langle   a^\dagger_0  + a_0  \rangle (t)$ refer to the photon and phonon displacements, stemming from Eq.~\eqref{eq_1}, while keeping the linear order of $\lambda$.

            \textbf{\blue{\textit{Results.}}}---Among various observables in the system, we focus on the occupations of subsystems; $n_{\rm pt}(t) = L^{-1} \langle b^\dagger b \rangle (t)$, $n_{\rm pn}(t) = L^{-1} \langle a^\dagger_0 a_0 \rangle (t)$, and $n_{\rm e}(t) = L^{-1} \sum_k \langle c^\dagger_k c_k \rangle (t)$ and derive corresponding equations of motion. To achieve sufficient accuracy of time propagation for a true NESS, we numerically solve the equations of motion up to a sufficiently long time, see Sec.~S1 of the Supplemental Material~(SM)~\cite{SM}.
            
            We consider a resonantly excited photon mode through laser interaction, setting $\omega = \eta$. Additionally, we resonantly couple phonons to the excited photon mode, ensuring that the laser field indirectly influences the phonons (i.e., we set $\omega_0 = \eta$). Previous studies on $\kappa$-salt~\cite{PhysRevX.10.031028,LeDe_2022,Eckhardt2022} demonstrated that the effective phonon frequency is $\omega_0 \approx 2 \,t_0$, positioned near the electronic band edge. As previously demonstrated~\cite{PhysRevB.108.L140305}, our mean-field-type approximation results in a minor deviation of the electron occupation from its equilibrium value of 1/2 in the cavity-free model, maintaining the model stability. This allows for the arbitrary setting of strengths for $g_{\rm q}$. In our system, the most important parameter is the effective phonon-photon coupling $\Delta$, which is responsible for transferring laser energy to the IR-active phonon, first, and second, to the electron. For the majority of parameters, we reach a NESS, as detailed in Sec.~S2 of the SM.\begin{figure}[t]
            	\centering
            	\setlength\abovecaptionskip{1pt}
            	\includegraphics[width=1\linewidth]{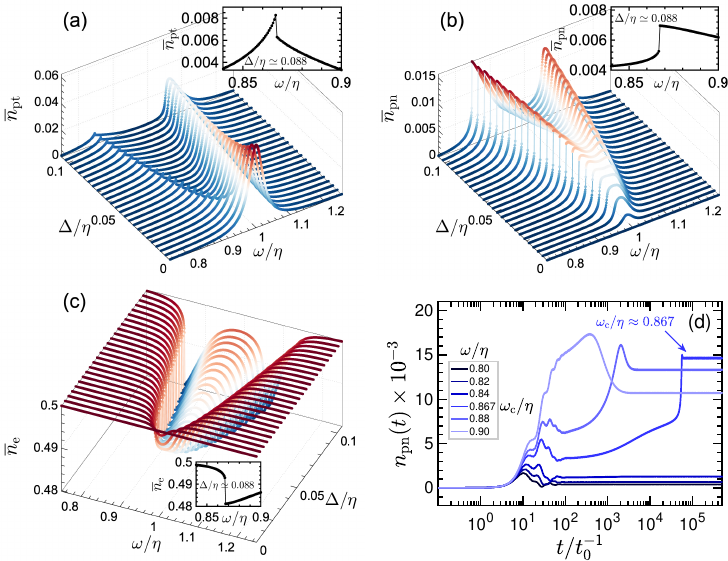}
            	\caption{\textbf{Asymmetrical nonequilibrium first-order phase transition.} The dynamics of occupied steady states for (a) photon, (b) phonon, and (c) electron are depicted across various phonon-photon couplings ($\Delta$) and driving frequencies ($\omega$). The laser amplitude is set to $\mathcal{A}_0/\gamma_{\rm pt} = 0.25$. Due to dissipation-induced nonlinearity in the relaxation process of electrons~\cite{PhysRevB.108.L140305}, leading to a reduction in the number of electrons in (c), shift and splitting of resonances in photon and phonon in (a) and (b) take place. This process facilitates the first-order phase transition, as highlighted in inset panels, and gives rise to the creation of hybridized states -- phonon-polaritons -- particularly for $\Delta > 0.05\,\eta$ values. The axes in (c) are swapped to increase the visibility of jumps at the phase transitions. Panel (d) shows the time evolution of dressed phonon occupation away from resonance for different driving frequencies at $\Delta/\eta = 0.088$, confirming the transition.}
            	\label{f3}
            \end{figure}
    
    		In addition to the provided data, Sec.~S3 of the SM offers supplementary information on various $g_{\rm q}$ and $\lambda$, confirming that the chosen values $(g_{\rm q}/\omega_0,\lambda) = (2,0.1)$ are proper choices following Ref.~\cite{PhysRevB.108.L140305}. Further, we proceed with weak damping rates for all degrees of freedom; throughout the paper, we fix $\gamma_{\rm pn}/\omega_0 = 0.05$, $\gamma_{\rm pt}/\eta = 0.05$, and $\gamma_{\rm e}/t_0 = 0.01$. While finite electron damping $\gamma_{\rm e} \neq 0$ is necessary to reflect the transient feedback effects between phonons/photons and electrons in the relaxation processes, its strength is not important. For further details on the effect of phenomenological photon and phonon damping rates, see Sec.~S4 of the SM. As for the laser amplitude, we set $\mathcal{A}_0/\gamma_{\rm pt} \ll 1$ to be weak enough to avoid system heating; we normalize it to the photon damping because a driven damped harmonic oscillator has an occupation of $(\mathcal{A}_0/\gamma_{\rm pt})^2$ in the decoupled phase, which we will use as a benchmark for the coupled phase~\cite{yarmohammadi2020dynamical}. Finally, to eliminate finite-size effects, we fix $L = 2001$ for the chain length. 

    	In what follows, we consider two scenarios: one without photon loss~($\kappa = 0$) from the cavity and one with photon loss~($\kappa \neq 0$). To investigate the properties of the NESS, we compute the average of temporal coherent signals over one period.

We start with the effect of $\Delta$ on occupations in the NESS in Fig.~\ref{f3}(a-c). Panel (d) specifically considers $\Delta/\eta = 0.088$. When there is no phonon-photon coupling ($\Delta = 0$), all entities peak at the resonance $\omega = \omega_0 = \eta$. However, once the photon interacts with the phonon and electron, resonances experience shifts and splittings, attributed to the formation of a ``phonon-polariton''. This observation aligns well with findings in Refs.~\cite{Eckhardt2022,Sentefeaau6969,LeDe_2022}. Nevertheless, the behavior of the lower polariton in photon occupation differs from that in phonon and electron occupations, resulting in an asymmetrical presence in the photon occupation. The primary reason for this distinction lies in the dressing effect in the model, where the electric field directly influences the photon occupation through $\widetilde{\mathcal{E}}(t) \approx {} \mathcal{E}(t) -\Delta \,p_{\rm pn}(t)$, where $p_{\rm pn}(t) = i\,L^{-1/2}  \langle   a^\dagger_0 - a_0  \rangle (t)$ refers to the phonon momentum. In contrast, the strength of this effect is comparatively weaker in the electron and photon occupations, where they are indirectly influenced by the laser field.

Remarkably, simultaneous with the formation of asymmetrical and symmetrical branches in photon and phonon/electron occupations during phonon-polariton development, a notable sharp jump is observed only for the lower polariton at various critical drive frequencies and $\Delta > 0.05 \,\eta$, as illustrated in the inset panels. The time evolution of the photon occupation in Fig.~\ref{f3}(d) further supports the occurrence of a phase transition at $\Delta/\eta = 0.088$ and $\omega/\eta \approx 0.867$. Indeed, at the critical drive frequency, the phonon occupation initially surges, reaching a quasi-divergence, but immediately descends to another NESS plateau (the highest value). The emergence of quasi-divergence signifies the first-order phase transition. This represents a novel theoretical finding regarding phonon-polaritons out of equilibrium that has not been uncovered previously.

By leveraging the dominant harmonics of oscillations in the NESS for all entities, we analytically confirm that the singularities in phonon displacement emerge at drive frequencies smaller than single-photon frequency, $\omega < \eta$, see Sec.~S5 of the SM. This arises from the solutions of a polynomial equation for $|q^{\rm pn}_1|$ (the first harmonic of $q_{\rm pn}(t)$), wherein its minimum reaches zero. From a physical standpoint, a zero minimum of phonon and photon displacements means that the objects are instantaneously frozen at the critical drive frequencies for $\Delta > 0.05 \,\eta$. The first-order phase transition is a direct consequence of such freezing-like effects.\begin{figure}[b]
	\centering
	\setlength\abovecaptionskip{1pt}
	\includegraphics[width=0.8\linewidth]{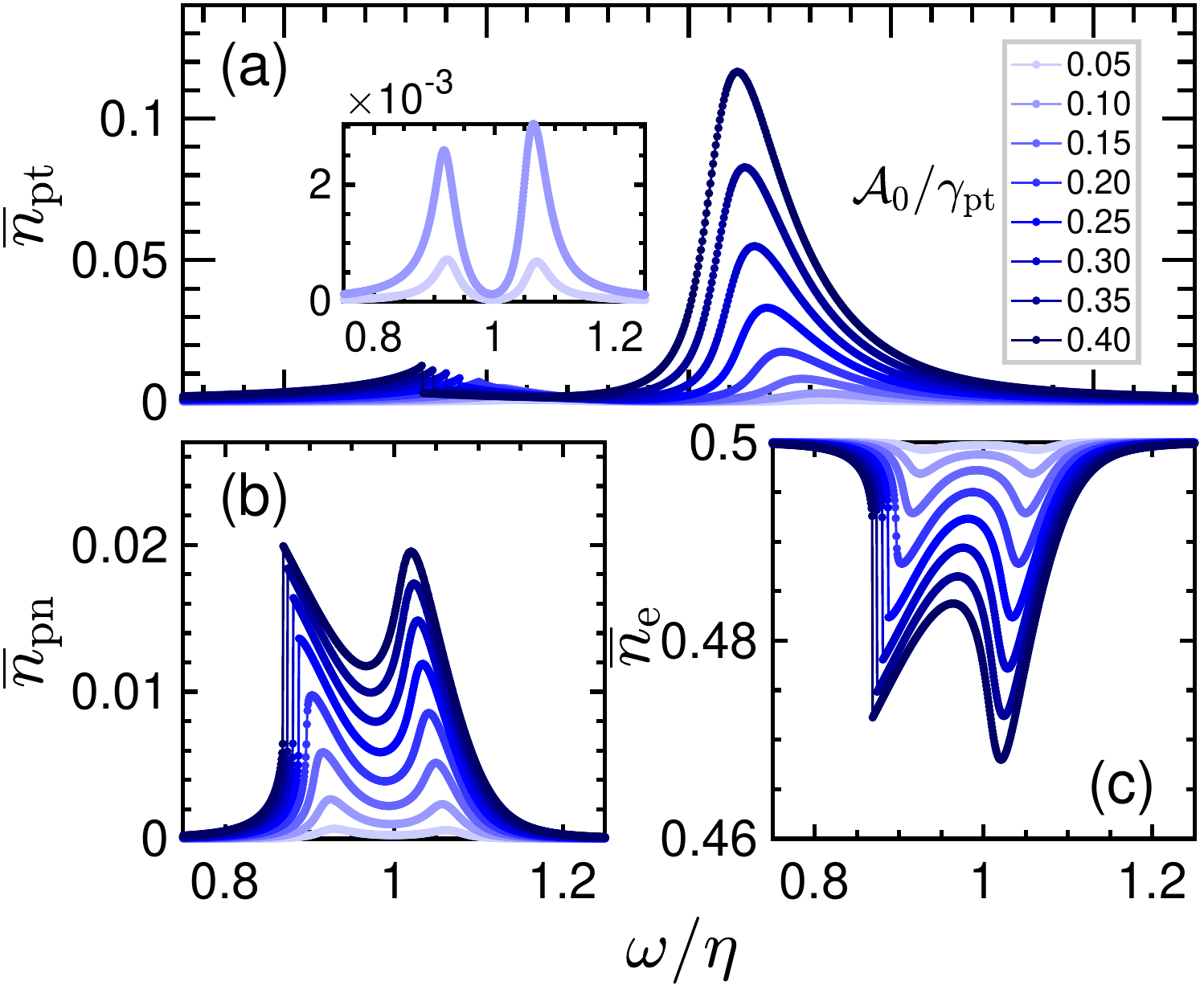}
	\caption{\textbf{Elimination of nonequilibrium first-order phase transition with weak laser field.} The role of laser amplitude on the first-order phase transition and the symmetrical responses of lower and upper polaritons in the NESS is illustrated for (a) photon, (b) phonon, and (c) electron at $\Delta/\eta \approx 0.1$. As $\mathcal{A}_0$ decreases, the lower and upper branches of the polariton display symmetry in the photon sector, and the first-order phase transition disappears in the entire system. This happens because of the reduced dressing effect caused by the laser field.}
	\label{f4}
\end{figure}

In the following analysis, we investigate the robustness of the phase transition and symmetrical responses of lower and upper polaritons, depicted in Fig.~\ref{f4}, under various laser amplitudes $\mathcal{A}_0$. In experimental settings, laser amplitude plays a crucial role in unveiling unique behaviors and potentially tracking phase transitions. As the laser amplitude decreases, the lower and upper branches of the phonon-polaritons symmetrically align around the resonance peak at $\omega/\eta = 1$, as seen in the inset panel of Fig.~\ref{f4}(a). The rationale behind this remains consistent with previous explanations; the weakening of the dressing effect with smaller $\mathcal{A}_0$ values causes the photon occupation to behave similarly to the phonon and electron. Moreover, we observe that such a weak drive field leads to the revocation of the first-order phase transition. This implies that the minimum values of phonon and photon displacements persist as finite without vanishing if the laser amplitude is sufficiently weak. This suggests that under such conditions, these objects never experience instantaneous freezing in the system, in contrast to the regime with strong $\mathcal{A}_0$.\begin{figure}[t]
			\centering
   \setlength\abovecaptionskip{1pt}
		\includegraphics[width=0.8\linewidth]{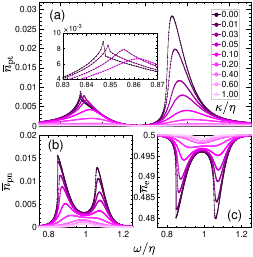}
			\caption{\textbf{Elimination of nonequilibrium first-order phase transition and phonon-polariton with strong photon loss.} The influence of photon loss $\kappa$ on the dynamical responses of phonon-polaritons in the NESS of the (a) photon, (b) phonon, and (c) electron at $\mathcal{A}_0/\gamma_{\rm pt} = 0.25$ and $\Delta/\eta \approx 0.1$. When photon loss $\kappa/\eta$ surpasses 0.03 and 0.2, the first-order phase transition and polaritons vanish, respectively, due to the blocking of input power from entering the cavity, as it is dissipated to the vacuum through photon loss.}
			\label{f5}
		\end{figure}
  
  Up to this point, we have considered a scenario where the system lacks photon loss from the cavity, leading to the formation of phonon-polaritons and the occurrence of the first-order phase transition under specific parameter conditions. For the sake of completeness, in Fig.~\ref{f5}, our focus turns on the effect of photon loss, addressing how the aforementioned dynamic features respond to changes in $\kappa$. In this scenario, we observe that for $\kappa/\eta > 0.03$, the first-order phase transition does not occur for our choice of remaining parameters. Additionally, the presence of the phonon-polariton is noticeable up to $\kappa/\eta \approx 0.2$, but beyond this threshold, polaritons fail to form. This can be understood by noting that with the inclusion of photon loss $\kappa$, a significant portion of the input laser power is dissipated to the vacuum before transferring energy to the single cavity mode and other components. Essentially, the resonance response of the system remains robust against couplings when exceeding $\kappa/\eta = 0.2$. 
  
           \textbf{ \blue{\textit{Experimental perspective.}}}---While the details of the couplings strongly depend on the actual material at hand, particularly the electron-phonon coupling, we can propose a physical realization of the first-order phase transition. Our findings indicate that the interaction between phonons and photons plays a key role in determining novel phenomena, motivating us to focus on the phonon as an easily accessible lattice degree of freedom in reality. By driving phonons through a terahertz field-driven optical photon, a portion of the absorbed laser power, transitioning from the excited phonon to the environment inside the cavity, can be described by $I(t) = \mathcal{M} \omega_0 \gamma_{\rm ph} n_{\rm ph}(t)$, where $\mathcal{M}$ represents the mole amount of thin film of our sample determined by its volume and molar density. To mitigate thermal effects, the sample can be placed in contact with a cold sink (characterized by temperature $T$, mass $m$, and a small heat capacity $C(T) \propto T$). With this, we can estimate the timescale at which the system maintains the NESS to observe phase transition, given by $t_{\rm exp} = m \int_0^T C(T) \, d\,T/\overline{I}$. For instance, consider a chain with hopping energy of $t_0 \approx 1$ THz, where the phonon frequency $\omega_0 \approx 2\,t_0$ in our simulations lies near the electronic band edge. For a chain with approximately $1\%$ phonon occupation per site~(see panel (b) of above figures), a cold sink possessing a mass of 2 g and a temperature of 2 K, the NESS can be maintained for up to $t_{\rm exp} \approx 500$ ns, providing sufficient time to observe the phase transition in the NESS. This is reachable using existing terahertz laser sources~\cite{Biasco2018, Khalatpour2021, Liebermeister2021,Yang2023}.
            
            \textbf{\blue{\textit{Summary and outlook.}}}---The occurrence of a nonequilibrium first-order phase transition stems from nonlinear dissipative dynamics in a driven fermion chain, where the electron density is coupled quadratically to an infrared-active phonon. Our current investigation unveiled that a driven optical cavity can finely adjust the ultrafast dynamics of particles in the non-equilibrium steady states (NESS) of such a chain. To analyze this system, we employed the Lindblad formalism, dipole approximation, and mean-field theory. Laser excitation of a single photon mode gives rise to hybridized states, particularly phonon-polaritons in the NESS primarily formed by the coupling between the driven photon and phonon.
            
   	 We determine that, in the absence of photon loss from the cavity, only the lower phonon-polariton experiences the first-order phase transition in the NESS, accompanied by the signature of the terahertz field-induced dressing effect in the (a)symmetrical responses of polaritons. Our findings suggest that a laser field lacking sufficient strength is incapable of triggering the first-order phase transition, but it does restore the symmetry of the lower and upper polaritons. Additionally, photon loss disrupts the phonon-polariton and its distinctive phase transition characteristic. These results bear significance for laser-driven switching and control processes in driven open quantum systems. 

    We briefly comment that a chirp drive field can effectively modulate all these dynamic features via hysteresis processes, enabling the manifestation of a phase transition for an alternative set of parameters as well~\cite{PhysRevB.108.L140305,yarmohammadi2023laserenhanced}. For example, it remains interesting to explore whether the chirp can prompt a first-order phase transition to the upper polariton of our current model. Moreover, the induction of a first-order phase transition in the presence of dissipation and the nonlinear coupling of electrons to photon-driven phonons in a cavity raises important questions. Specifically, it prompts inquiries about the cavity-mediated induction of phonon-induced interactions between electrons with spin degrees of freedom for real materials in the cavity. For example, it will be interesting to study the modalities afforded by a cavity field in the presence of dissipation for the induction of coherent phase transitions of the electrons such as cavity-mediated superconductivity and density waves~\cite{PhysRevB.99.020504,PhysRevLett.122.167002,Ruggenthaler2023}. We will address these questions in future work. 
    
	\textbf{\blue{\textit{Acknowledgments.}}}---We greatly thank Edoardo Baldini for helpful discussions. This work was performed with support from the National Science Foundation (NSF) through award numbers MPS-2228725 and DMR-1945529. Part of this work was performed at the Aspen Center for Physics, which is supported by NSF grant No. PHY-1607611. Funded by the European Union (ERC, QuSimCtrl, 101113633). Views and opinions expressed are however those of the authors only and do not necessarily reflect those of the European Union or the European Research Council Executive Agency. Neither the European Union nor the granting authority can be held responsible for them.

}
	\bibliography{bib}
\end{document}